\documentclass[sigconf]{acmart}

\usepackage[utf8]{inputenc}
\usepackage[T1]{fontenc}

\usepackage{xcolor}
\usepackage[utf8]{inputenc}
\usepackage[T1]{fontenc}
\usepackage{graphicx}
\usepackage{subcaption}
\usepackage[export]{adjustbox}
\usepackage{textcomp}
\usepackage[hang,flushmargin]{footmisc}
\usepackage{url}
\usepackage{import}
\usepackage{color}

\usepackage{soul}
\usepackage{pifont}
\usepackage{enumitem}
\usepackage{multirow}
\usepackage{blindtext}
\usepackage{dirtytalk} 
\usepackage{tabularx} 
\usepackage{multirow}
\usepackage{listings} 
\usepackage{todonotes}
\usepackage{mathtools}
\usepackage{cleveref}
\usepackage{rotating}

\graphicspath{ {./Figures/} }
\usepackage[autostyle=true]{csquotes}
\MakeOuterQuote{"}

\lstdefinestyle{python}{ 
    xleftmargin=6.0ex,
    xrightmargin=2.0ex,
    numbers=left,
    frame=single
}
\definecolor{babyblue}{rgb}{0.63, 0.79, 0.95}
\definecolor{bubblegum}{rgb}{0.99, 0.76, 0.8}
\definecolor{aliceblue}{rgb}{0.94, 0.97, 1.0}
\definecolor{lightsalmon}{rgb}{1.0, 0.63, 0.48}
\definecolor{emerald}{rgb}{0.31, 0.78, 0.47}
\definecolor{grannysmithapple}{rgb}{0.66, 0.89, 0.63}
\definecolor{moccasin}{rgb}{0.98, 0.92, 0.84}
\definecolor{naplesyellow}{rgb}{0.98, 0.85, 0.37}
\definecolor{lightcyan}{rgb}{0.88, 1.0, 1.0}
\definecolor{lightgoldenrodyellow}{rgb}{0.98, 0.98, 0.82}
\definecolor{darktangerine}{rgb}{1.0, 0.66, 0.07}

\usepackage{color}

\RequirePackage{fontawesome}

\definecolor{gray50}{gray}{.5}
\definecolor{gray40}{gray}{.6}
\definecolor{gray30}{gray}{.7}
\definecolor{gray20}{gray}{.8}
\definecolor{gray10}{gray}{.9}
\definecolor{gray05}{gray}{.95}
\definecolor{green}{rgb}{0.0, 0.5, 0.0}
\definecolor{arsenic}{rgb}{0.23, 0.27, 0.29}
\definecolor{yellow}{rgb}{1.0, 0.75, 0.0}

\newlength\Linewidth
\def\findlength{\setlength\Linewidth\linewidth
    \addtolength\Linewidth{-4\fboxrule}
    \addtolength\Linewidth{-3\fboxsep}
}

\newenvironment{rqbox}{\par\begingroup
	\setlength{\fboxsep}{5pt}\findlength
	\setbox0=\vbox\bgroup\noindent
	\hsize=0.95\linewidth
	\begin{minipage}{0.95\linewidth}\normalsize}
	{\end{minipage}\egroup
	\textcolor{gray20}{\fboxsep1.5pt\fbox
		{\fboxsep5pt\colorbox{gray05}{\normalcolor\box0}}}
	\endgroup\par\noindent
	\normalcolor\ignorespacesafterend}

\newcommand{\eg}{\emph{e.g.,}\xspace}

\begin{document}

\title{On the Empirical Evidence of Microservice Logical Coupling}
\subtitle{A Registered Report}

\author{Dario Amoroso d'Aragona}
   \affiliation{
   \institution{Tampere University\\} 
   \city{Tampere}
   \country{Finland}
 }
 \email{dario.amorosodaragona@tuni.fi}

\author{Luca Pascarella}
   \affiliation{
   \institution{ETH Zurich\\} 
   \city{Zurich}
   \country{Swizerland}
 }
 \email{lpascarella@ethz.ch}

 \author{Andrea Janes}
 \affiliation{
   \institution{FHV Vorarlberg University of Applied Sciences\\} 
   \city{Dornbirn}
   \country{Austria}
 }
 \email{andrea.janes@fhv.at}

 \author{Valentina Lenarduzzi}
 \affiliation{
   \institution{University of Oulu\\} 
   \city{Oulu}
   \country{Finland}
 }
 \email{valentina.lenarduzzi@oulu.fi}

 \author{Rafael Pe\~naloza}
 \affiliation{
   \institution{University of Milano-Bicocca\\} 
   \city{Milan}
   \country{Italy}
 }
 \email{rafael.penaloza@unimib.it}

 \author{Davide Taibi}
 \affiliation{
   \institution{University of Oulu\\ Tampere University} 
   \city{Oulu}
   \country{Finland}
 }
 \email{davide.taibi@oulu.fi}


\renewcommand{\shortauthors}{Amoroso d'Aragona, et al.}


\begin{abstract}
[Context] Coupling is a widely discussed metric by software engineers while developing complex software systems, often referred to as a crucial factor and symptom of a poor or good design. Nevertheless, measuring the logical coupling among microservices and analyzing the interactions between services is non-trivial because it demands runtime information in the form of log files, which are not always accessible.

[Objective and Method] In this work, we propose the design of a study aimed at empirically validating the Microservice Logical Coupling (MLC) metric presented in our previous study. In particular, we plan to empirically study Open Source Systems (OSS) built using a microservice architecture. 

[Results] The result of this work aims at corroborating the effectiveness and validity of the MLC metric. Thus, we will gather empirical evidence and develop a methodology to analyze and support the claims regarding the MLC metric. Furthermore, we establish its usefulness in evaluating and understanding the logical coupling among microservices.



\end{abstract}


\maketitle

\section{Introduction}
Coupling and cohesion are two fundamental metrics in software engineering: coupling measures the degree of interdependence between modules (low is preferred) and the cohesion of a module indicates the extent to which its individual components are needed to perform its task (high is preferred)~\cite{Fenton1997}. Low coupling and high cohesion are often mentioned in connection with maintainable code: code with low coupling means that one module can be modified without impacting other modules; a highly cohesive module means that that module has a single, well-defined purpose. 

This paper studies to what extent software systems whose development architecture is based on microservices respect a \emph{``low coupling''}.
In Microservice Architecture (MSA), a system is composed of independently deployable microservices where each service has 
a single responsibility, meaning they only manage one specific part of the organization's needs~\cite{Fowler2004}. 
Microservices should be loosely coupled since coupling increases maintenance effort and---particularly in the context of microservices---increases the need for synchronization between teams. A development team in charge of a single service should need to know as little as possible about other services. Highly coupled services might require synchronizing with other services before deploying new features, reducing the benefits of microservices. Coupling not only slows down the development process but also impacts other qualities e.g., performance, since communication between services is slow compared to communication within a service. Along the same lines, microservices should be highly cohesive, keeping all the related logic in a single service rather than splitting it into multiple services~\cite{Newman2021}.

Software engineers agree on the importance of low coupling between microservices referring to ``independence between teams''~\cite{Hastie2022}, ``independent deployment''~\cite{Ford2021} or ``no need to synchronize between teams before deploying''~\cite{Richardson2022}. 
Indeed, practitioners often mention low coupling as a main benefit of microservices \cite{TaibiIEEECloud2017,Soldani2018}.

Although practitioners agree about the impact of these metrics on the software quality, they remain reluctant to a practical usage by choosing a short-circuit approach based on their gut feeling to structure a system into services, causing an uncontrolled degree of coupling. 
Possible reasons for not adopting the  metrics proposed by the scientific community are over-optimism, inadvertence, or the unavailability of data. As an example, \cite{Taibi2020, Gadler2017} assume the availability of log files describing calls to software components, used to extract usage processes, which are then used to propose microservices. Such logs are not always available or might require major changes in the system under development. Moreover, the measurement of coupling proposed in the literature is based on the static~\cite{cerny22} or dynamic analysis~\cite{Gortney22} of source code. Coupling between teams such as the need to wait or to synchronize with other service teams before committing is not captured by such metrics. 
Previous works addressed the problem of coupling in monolithic systems thoroughly. Also, the concept of \textit{logical} coupling was introduced, i.e., coupling, which is not based on a dependency in source code, but on the implicit dependency between artifacts that are often changed together. In particular, \cite{Ambros2009} extended the logical coupling metric, originally proposed by \cite{Robbes2008} to capture whether changes made in a predefined time window are logically coupled. 

In this research, our objective is to propose a study design aimed at empirically validating the Microservice Logical Coupling (MLC) metric we introduced in our previous work~\cite{Amoroso2023}. To this aim, we plan to conduct an empirical study by focusing our analysis on Open Source Systems (OSS) whose software design is based on microservices architecture. The primary focus of this study is to provide empirical evidence supporting the effectiveness and validity of the Microservice Logical Coupling (MLC) metric. 

The expected outcome from this work is two-fold: 

\begin{itemize}
    \item a new validated context-aware metric to calculate the logical coupling between microservices;
    \item a detailed replication package reporting all the data and scripts to open further research studies. 
\end{itemize}

Practitioners can benefit from these results from both research and practical perspective by assessing one aspect of the coupling between microservices. 

\textbf{Paper structure:} Section~\ref{sec:Background} describes the background, Section~\ref{sec:Relatedwork} depicts related work,  Section~\ref{sec:EmpiricalStudy} introduces our study design, Section~\ref{sec:Execution} reports the study execution, Section~\ref{sec:ThreatsValidity} discusses the possible threats, Section~\ref{sec:RiskManagement} shows the risk management, and Section~\ref{sec:Conclusion} draws conclusions and future work.

\section{Background}
\label{sec:Background}

Complex software systems are not always created by ardent practitioners for various reasons. Developers are continuously dealing with project demand and restrictions. As a consequence, developers incur the potential risk of performing changes to the software which aren't related to their expertise, assignments, or component.
Logical coupling uses a software system's development history to detect change patterns among code units that are modified together in order to spot entangled changes in versioning systems.

Robbes et al.'s~\cite{Robbes2008} metric, adopted by D'Ambros et al.~\cite{Ambros2009}, measures whether changes performed within a specified time range are logically coupled. \\
In our previous work ~\cite{Amoroso2023}, we  investigated changes involving multiple microservices committed atomically in the same working unit. 
In their work, they consider two microservices logically coupled if these have been co-changed more than 5 times during the history of the project.
However, the proposed metric~\cite{Amoroso2023} considers neither the decoupling of two microservices -- thus when due to some refactoring in the code the two microservices are not logically coupled anymore -- nor the evolution of the coupling during the development process nor the context of the project -- commit frequency, number of developers, number of microservices and so on -- that can influence the metric leading in an inaccurate result. 
For this reason, we proposed a new metric that can be used to analyze the evolution of the coupling during the history of the project, allowing us to study when two microservices start to be logically coupled and when they could be considered decoupled. 
We will perform different analyses to understand how we can define a threshold to consider two microservice logically coupled based on a single specific project, in this way, we can develop a proposal to identify the best threshold depending on the specified project.
Furthermore, we will perform a study to understand which factors can influence the metrics and how we can mitigate this influence by setting differently the metrics parameters.

\section{Related work}
\label{sec:Relatedwork}
Several metrics have been suggested for monolithic systems, some of which have been adapted for service-based systems, particularly for microservices, as demonstrated in previous research (see e.g.,~\cite{Fenton1997}). Bogner et al. \cite{Bogner2017} conducted a systematic literature review on maintenance metrics for microservices, with a focus on service-based systems rather than metrics developed for object-oriented systems. Their findings indicate that most of the metrics that were originally designed for monolithic systems and Service Oriented Architectures are also relevant in the context of microservices. This study is considered to be the first to focus exclusively on microservices, and subsequent research has expanded on these findings to further enhance our understanding of the topic.

\textcolor{black}{Apolinaro et al.~\cite{Apolinaro} presented a theoretical use case proposing a roadmap to apply four metrics defined by~\cite{Bogner2017}:}
\begin{itemize}
    \item \textcolor{black}{Absolute Importance of the Service (AIS): number of consumers invoking at least one operation from a service}
    \item \textcolor{black}{Absolute Dependence of the Service (ADS): number of services on which the service depends
    \item Service Coupling Factor (SCF) as the density of a graph’s connectivity. SCF=SC/N$^2$-N where SC is the sum of  calls between services, and N is the total number of services.
    \item Average Number of Directly Connected Services (ADCS): the average ADS metric of all services.}
\end{itemize}


In their study, Taibi et al. \cite{Taibi2020} suggested four measures (coupling between microservices, number of classes per microservice, number of duplicated classes, and frequency of external calls) to be considered when breaking down an object-oriented monolithic system into microservices. However, these metrics rely on the concept of classes and lack empirical validation. On the other hand, Panichella et al. \cite{Panichella2020} proposed a structural coupling metric, which was tested in 17 open-source microservice-based projects (Imranur et al.\cite{Imranur2019}). The metric calculates the coupling of services at runtime based on the inbound and outbound calls between services.

In our recent work~\cite{Amoroso2023}, we proposed a new metric to measure the level of logical coupling between microservices, which is based on the analysis of commits to versioning systems. To validate the effectiveness of this metric, the authors collected data from 145 open-source microservices projects and performed an initial analysis. The results show that logical coupling has a significant impact on the overall system and tends to increase over time. 




\section{The Empirical Study}
\label{sec:EmpiricalStudy}
We now describe our empirical study reporting goal, research questions, context, data collection, and data analysis. 

\subsection{Goal and Research Questions}
\label{sec:ResearchQuestions}
Our goal is to empirically validate Microservice Logical Coupling (MLC) metric we defined in our previous work~\cite{Amoroso2023}. 
%
Therefore, we formulated the following three Research Questions (RQs):

\begin{center}	
	\begin{rqbox}
		\textbf{RQ$_1$.} \emph{Does the MLC metric respect the representation condition of measurement?} 
  \end{rqbox}	 
\end{center}

In this RQ, we will consider measurement as a mapping from the empirical world to the formal, relational world. Consequently, a \textit{measure} is the number or symbol assigned to an entity by this mapping in order to characterize an attribute \cite{Fenton2014}. Now for us, the empirical relationship we want to measure is logical coupling in the context of microservices.  

That is, the representation condition asserts that a measurement mapping (a measurement) M must map entities into numbers and empirical relations into numerical relations in such a way that the empirical relations preserve and are preserved by the numerical relations.
For example, if we measure height using a meter, the measurement method  M that we use to determine height needs to respect the following representation condition: A is taller than B if and only if M(A)>M(B) \cite{Fenton1994,Fenton2014}. 

In this context, ``validating a software measure is the process of ensuring that the measure is a proper numerical characterization of the claimed attribute by showing that the representation condition is satisfied \cite{Fenton1997}'', which is what RQ$_1$ wants to investigate.

\begin{center}	
	\begin{rqbox}
		\textbf{RQ$_2$.} \emph{How does the project context impact the MLC metric?}  
  \end{rqbox}	 
\end{center}


When validating a measure, it ``must be viewed in the context in which it will be used'' \cite{Fenton1997}. Therefore, validation must take into account the measurement's purpose since a measure may be valid for some uses but not for others \cite{Fenton1997}. Therefore, with RQ$_2$, we want to investigate how the following factors (F) influence the accuracy of the metric:


\begin{itemize}
    \item[F1:] \textit{Commit scope} Since we assume that a commit indicates a change in a particular microservice, large refactoring or application-wide user interface changes should be excluded from this consideration to avoid them from being interpreted as that all microservices are coupled.

    \item[F2:] \textit{Commit size (F2)} It is necessary to study in detail which commits need to be excluded or particularly considered in the proposed metric. Moreover, the more rarely a developer commits while changing code, the more changes are included in each individual commit, and thus this increases the probability that two different microservices seem coupled. For example, in the extreme case of one commit per month, all microservices might appear coupled, so it is necessary to understand the sensitivity of our metric with respect to the commit size. 

    \item[F3:] \textit{Microservice size (F3)} It is common practice to suggest small changes to facilitate the code review process, and to reduce the need for regression testing. For example, a study showed that in open source projects a change in the 80\% of the time involves four files \cite{Arafat2009}. 
So if a change in 80\% of cases involves 4 to 5 files, it seems more likely that in the extreme case of having very small microservices will result in a higher probability of logical coupling. Conversely, having large microservices this consideration seems less likely. 

\end{itemize}

\subsection{Data Collection}

To understand the coupling and decoupling of microservices, we need to know when different microservices are updated, and how often they are modified in correspondence to each other. We will collect all the commits in a project, grouped by commit day, and by microservice. That is, for each day, we will signal which microservices were updated.

To avoid spurious conclusions derived from gaps in the work
, we will remove from the dataset all days in which no commits were made to the project. Hence we consider only ``active'' days (rather than calendar days).

\subsection{Data Analysis}

    
    
\begin{figure}[h]%
\centering
\includegraphics[width=1\linewidth, trim={0 0 0 0},clip]{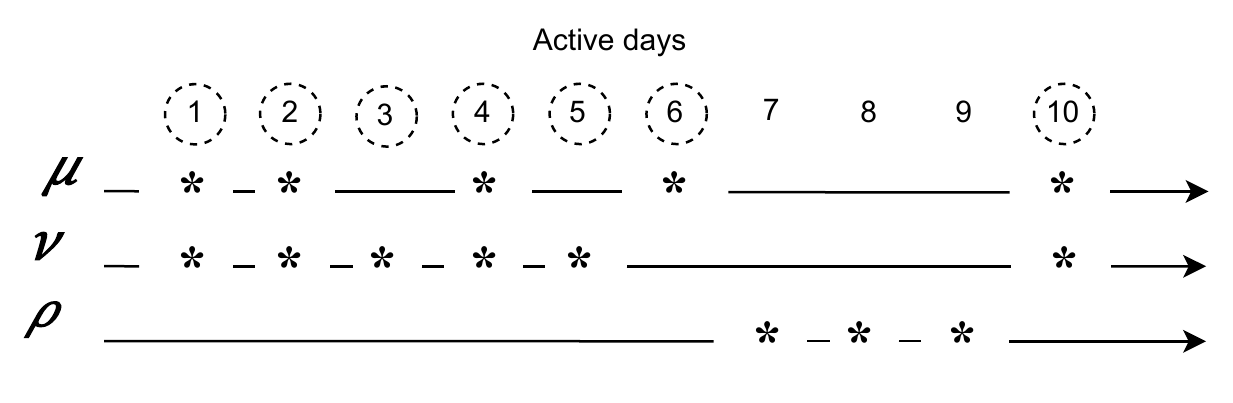}
\caption{Selection of the relevant active days for the microservices couple $(\mu,\nu)$ }
\vspace{-5mm}
\label{fig:active_days}
\vspace{0.5cm}
\end{figure}

\begin{figure}[b]
    \centering
    \includegraphics[width=1\linewidth, trim={0 0 0 0},clip]{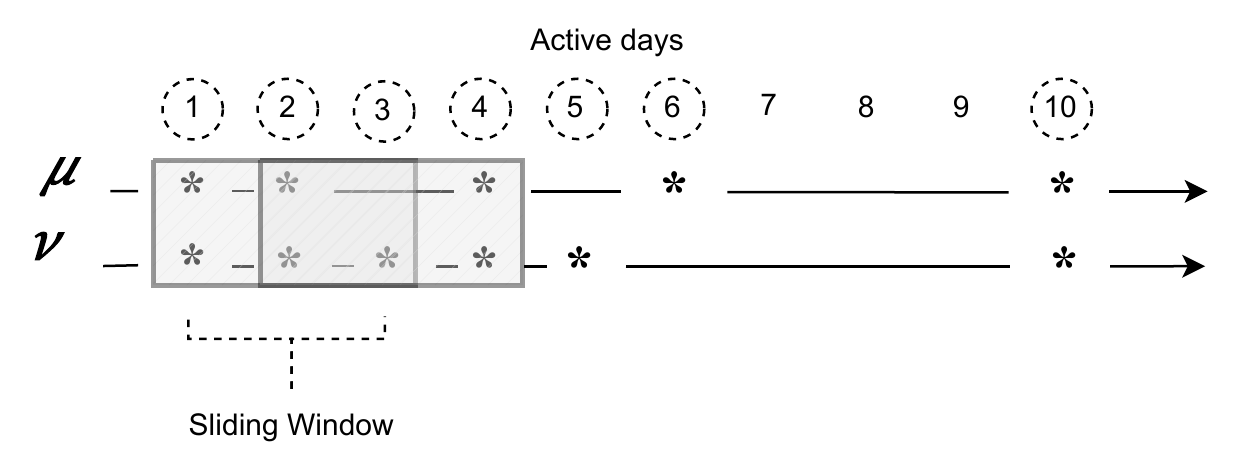}
    \caption{Three days sliding window example}
    \vspace{-5mm}
    \label{fig:sliding window}
    \vspace{0.5cm}
\end{figure}

In order to answer our RQ$_1$, we need to measure the coupling of microservices and their evolution over time. We will use a \emph{sliding window} approach. Fixed a number $n$, we look at each window of $n$ consecutive (active) days. For each pair $\mu,\nu$ of microservices of interest, we will count the number of instances in which \emph{both} services were updated within the window. Specifically, a microservice is updated if at least one of its associated files is modified at the commit. The proportion of days with shared updates is associated with the \emph{coupling value} at the end of the window. For example, if $\mu$ is updated on days $\{1,2,4,6,10\}$ and $\nu$ is updated on active days $\{1,2,3,4,5,10\}$ (\Cref{fig:active_days}), assuming a sliding window of 3 days, during the first window the microservices are mutually updated in two (1 and 2) of the three days (\Cref{fig:sliding window}), yielding a coupling value of $2/3$. 
The second window considers days 2--4, and yields a coupling value of $2/3$ again. On the third window (days 3--5) the coupling value decreases to $1/3$ and remains like that in the successive window.
The days 7, 8, and 9 are not considered even if they are active days---$\rho$ is committed---because they are not relevant for the couple $(\mu,\nu)$.
We consider only the number of days (in the window) when at least one of the services is updated. In the example above, the days 7--9 are removed since neither $\mu$ nor $\nu$ is updated on those
days (they are thus irrelevant for the pair). Note that, thanks to this focus on relevant active days, the coupling value between the services remains at least $1/3$, in contrast with the 0 that
is observed if days 7--9 were included.
In this way, we can model the decrease in coupling. If we considered all the active days, we would have a value of 0, of occurrences in a single day, both when none of the microservices is committed and when only one of the two is committed.
But active days when neither of the two microservices of interest is committed add no information and act as noise in the data.
Furthermore, if two microservices are seldom updated, their coupling value would be small, even if they are constantly updated together. 

The coupling value increases as $\mu$ and $\nu$ co-occur more often, and decreases otherwise. Using these values, we can apply different techniques for understanding the evolution of coupling between the services. Given a threshold, we can find instances where the services are ``sufficiently'' coupled. But using specific numerical values allows us to apply time-series analyses---in particular, time-series segmentation---to understand the periods of time where the coupling is growing, stable, or decreasing, respectively.

The size $n$ of the sliding window cannot be fixed \emph{a-priori}, as it depends on the specific properties of the project. Yet, it can be seen that a longer window yields a more stable behavior, where the numerical values differ less between successive active days. We will experiment with different window sizes (i.e., 10--30--100) to better understand this behavior.
Finally, rather than checking the co-occurrence of microservice updates, we can verify for \emph{logical} dependencies: if service $\mu$ is updated, do we expect to update $\nu$ as well? This value is obtained by modifying the proportion described before to the number of co-occurrences divided by the active days where $\mu$ is updated within the window. This can be seen as the conditional probability of modifying $\nu$ given that $\mu$ was modified. 
Importantly, contrary to ``pure'' coupling, this measure is not symmetric; that is, the likelihood of modifying $\nu$ when $\mu$ is changed is not necessarily equal to that of modifying $\mu$ under the assumption that $\nu$ is modified. This asymmetry allows us to better understand the relationships and dependencies between microservices.

Regarding RQ$_2$, to understand the influence of the three relevant factors on our metrics, we will analyze the impact of selecting different thresholds for filtering out commits, and for filtering microservices of different sizes, in the overall time series through a correlation test. 

\subsection{Verifiability and Replicability}
We will make the scripts and raw data available in our online appendix to allow verifiability and replicability.



\section{Execution Plan}
\label{sec:Execution}


This Section shows the execution plan we will aim to implement according to the study design defined in Section~\ref{sec:EmpiricalStudy}. The execution plan is constructed by splitting the activities into six main steps: project selection, retrieving co-changed microservices, time series analysis, conditional probability, exploratory analysis, and validation.

\subsection{Project Selection}

As the first step of our study, we must collect data from well-developed software projects. For this aim, we will initially focus on projects whose source code is systematically developed by using \textsc{git} versioning system and is hosted publicly on \textsc{GitHub}.
Then,  
we will define all the characteristics needed by a project to be included in our analysis (e.g., \emph{age, number of commits, number of developers. number of stars, number of watchers, supported by companies/governments/foundation}).
In addition, we will 
include only those projects that implement a consistent architecture based on microservices
.


\subsection{Retrieving co-changed microservices}
As a successive step, we will need to collect all microservices whose associated source code is modified within a single time window. More precisely, we refer to co-changed microservices as cases in which at least two files belonging to two different microservices are changed simultaneously, or in other words, they are changed within the same day. To do so, we will first focus on relevant active days by excluding all calendar days which do not present changes on the given microservices and then will extract all the files that changed on these days. Successively, we will create a map to associate each microservice with the related changed file. 

\subsection{Time series analysis}

In the third step, we will proceed with a time series analysis by considering all changes associated with each microservices. In particular, for each defined sliding window, we will count in how many active days, in the same sliding window, the microservices have been committed together divided by the length of the sliding window, in \Cref{fig:sliding window} the result of this step will be: $2/3$ for the first and second window and $1/3$ for the last three windows. 

\subsection{Conditional Probability}

In the fourth step, we will focus on the conditional probability. For each pair of microservices, we will extract the active days where at least one of the two microservices has been changed. It is worth noticing that the conditional probability of \Cref{fig:sliding window} depends on the order in which the two microservices are analyzed. For example, the conditional probability $P(\mu  \mid \nu)$ corresponds to $2/3$ while the conditional probability $P(\nu  \mid \mu)$ is $2/2$.

\subsection{Exploratory Analysis}

In the fifth step, we will extract the commit scope (\eg \emph{refactoring, bug fix, improvement, new feature}) by parsing the commit message.  We will extract the commit size (e.g \emph{chunks}). We will study how the results change using different \emph{thresholds} to consider two microservices logically coupled, different lengths of the \emph{sliding window} and filtering out the commit based on the commit scope and the commit size. We will compare the different results by grouping the projects by the number of microservices. Each variation produces a different time series. These variations will be compared through a correlation analysis.

\subsection{Validation}
In the last step, we will validate our metric. 
In particular, since we are interested in understanding (i) how the proposed metric depicts the evolution of the logical coupling over time and (ii) to what extent microservices labeled as coupled are effectively united, we will:
\begin{itemize}
    \item Manually calculate the coupling of a statistically relevant sample of microservices (e.g., confidence level 95\%, confidence interval 1) so as to create our ground truth. 
    \item Use accuracy measures (e.g., \emph{precision, recall, and f-measure}), checking whether the microservices labeled as coupled are actually coupled;
    \item Use the ground truth to evaluate the conditional probability and thus if our probability corresponds to a real update in real life;
    \item Check if the trend caught by our metric correspond to the actual trend.
\end{itemize}

\section{Threats to Validity}
\label{sec:ThreatsValidity}
In this section, we discuss the threats that might affect the validity of our empirical study.

We acknowledge the potential limitations of our research. Firstly, while we have chosen a diverse and up-to-date dataset, it may not encompass the entire spectrum of open-source microservices projects. Nonetheless, we believe it represents a valuable collection in terms of microservice quantity, project age, and developer involvement. Secondly, it's important to note that industrial projects might have different perspectives on coupling, particularly due to the emphasis placed on minimizing coupling in practitioner discussions. Nevertheless, exploring the microservice lifecycle (MLC) of a project can shed light on ineffective team dynamics and workflow structures, enabling organizational improvements.

Furthermore, our MLC validation analysis solely focuses on co-changes of microservices within the same commit. Consequently, it does not capture subsequent changes arising from the necessity to synchronize services. To address this, one potential solution could involve introducing a time window to account for co-changes. Additionally, considering the issues reported in the project's issue trackers could provide insights into whether developers are requesting synchronization with other services for their changes.

\section{Risk management} \label{sec:RiskManagement}

During the execution of this study, we foresee mainly two risks that might impact the results: \textit{Lack of OSS Projects} and the \textit{Unavailability of some of the original authors during the execution of the study}.

For the first aspect, there is a risk of not finding  OSS projects fulfilling our criteria. However, we are aware that different works have already performed studies investigating the quality of OSS Microservices, thus minimizing this risk. For the second aspect, since not all authors have permanent contracts, therefore there is the risk that some authors might change affiliations or not be available to continue this work actively. To mitigate this risk, the  authors committed to continuing this work even without some of the author's contributions.

\section{Conclusion}
\label{sec:Conclusion}
In this work, we describe the design of an empirical study aimed at understanding if the Microservice Logical Coupling (MLC) metric respect the representation condition of
measurement and how the project context impacts the MLC metric itself. 

We plan to shed light on the MLC,  executing this study on Open Source Projects developed with microservices. 

The execution of this study will allow us to understand if the MLC is a viable metric to measure logical coupling between microservices, and will allow researchers and practitioners to eventually use and extend it in future works.

\bibliographystyle{ACM-Reference-Format}

\bibliography{bibliography.bib}

\end{document}